\documentclass[]{spie}  

\usepackage{amsmath,amsfonts,amssymb}
\usepackage{graphicx}
\usepackage[colorlinks=true, allcolors=blue]{hyperref}

\newcommand{\ket}[1]{\ensuremath{\left|\right.\!{#1}\!\left.\right\rangle}}
\newcommand{\bra}[1]{\ensuremath{\left\langle\right.\!{#1}\!\left.\right|}}

\title{Hitting time for quantum walks of identical particles}

\author[a,b]{Alexey A. Melnikov}
\author[b]{Aleksandr P. Alodjants}
\author[a,c,d]{Leonid E. Fedichkin}
\affil[a]{Institute of Physics and Technology, Russian Academy of Sciences\\\newline Nakhimovskii prospekt 36/1, 117218 Moscow, Russia}
\affil[b]{ITMO University, Kronverksky prospekt 49, 197101 St. Petersburg, Russia}
\affil[c]{NIX, Zvezdniy boulevard 19, 129085 Moscow, Russia}
\affil[d]{Moscow Institute of Physics and Technology, Institutskii pereulok 9\\\newline 141700 Dolgoprudny, Moscow Region, Russia}

\authorinfo{Further author information: (Send correspondence to A.A.M.)\\A.A.M.: E-mail: melnikov@phystech.edu\\ A.P.A.: E-mail: alexander\_ap@list.ru \\ L.E.F.: E-mail: leonid@phystech.edu}

\begin{document}
\maketitle

\begin{abstract}
Quantum particles are known to be faster than classical when they propagate stochastically on certain graphs. A time needed for a particle to reach a target node on a distance, the hitting time, can be exponentially less for quantum walks than for classical random walks. It is however not known how fast would interacting quantum particles propagate on different graphs. Here we present our results on hitting times for quantum walks of identical particles on cycle graphs, and relate the results to our previous findings on the usefulness of identical interacting particles in quantum information theory. We observe that interacting fermions traverse cycle graphs faster than non-interacting fermions. We show that the rate of propagation is related to fermionic entanglement: interacting fermions keep traversing the cycle graph as long as their entanglement grows. Our results demonstrate the role of entanglement in quantum particles propagation. These results are of importance for understanding quantum transport properties of identical particles.
\end{abstract}


\section{INTRODUCTION}
\label{sec:intro}

The concept of random walks of quantum particles is defined by analogy to random walks of classical particles~\cite{PhysRevA.48.1687,Aharonov:2001:QWG:380752.380758}. Due to the superposition property of quantum particles, their dynamics differs fundamentally from classical particles.
This difference led to the development of the field of quantum walks~\cite{doi:10.1080/00107151031000110776,venegas2008quantum,venegas2012quantum} together with several algorithmic applications~\cite{Childs:2003:EAS:780542.780552,doi:10.1142/S0219749903000383,Ambainis:2007:QWA:1328722.1328730,Childs1,doi:10.1137/090745854,Childs2,portugal2013quantum}.
It was shown that on many graphs quantum walks of particles are faster than corresponding classical random walks~\cite{Magniez2012}. For some graphs the speed-up in hitting time is exponential~\cite{kempe2005discrete,PhysRevA.73.032341} and is polynomial for, e.g., lines and cycles, as also shown in Refs.~\citenum{solenov2006continuous} and~\citenum{fedichkin2006mixing}.

Quantum walks can be used to develop new tools in quantum information theory. These tools include, e.g., new algorithms for quantum computation and new quantum-enhanced machine learning methods.
As shown in Ref.~\citenum{melnikov2016quantum}, the new tools for quantum computing can be developed using two indistinguishable quantum particles in a cycle graph. It was shown that one can define qudits ($d$-dimensional quantum systems) by splitting a graph into two subgraphs~\cite{melnikov2016quantum,melnikov2014quantum,melnikov2016continuous}. If these two particles interact, then the corresponding qudits become entangled. Moreover, using different sizes of a cycle graph one can obtain a diverse structure of two-particle fermionic entanglement of high dimensions~\cite{melnikov2016continuous}. In Refs.~\citenum{melnikov2017entanglement,melnikov2018fermionic} the periodicity of this fermionic entanglement dynamics was demonstrated for quantum walks on $n$-cycle graphs.

In this paper, we study hitting times for continuous-time quantum walks of two identical particles on a cycle graph. The remainder of this paper is structured in the following way. We first describe the basic notions of continuous-time quantum walks and continuous-time random walks in Section~\ref{sec:results}. After this description, we discuss and compare hitting times of quantum and classical particles. We next consider two identical particles and study their hitting times on cycle graphs of different sizes. Then, we summarize the results of the paper in Section~\ref{sec:conclusion} and discuss possible applications of these results.

\section{Results}
\label{sec:results}

\begin{figure}[t!]
   \begin{center}
   \includegraphics[width=1\linewidth]{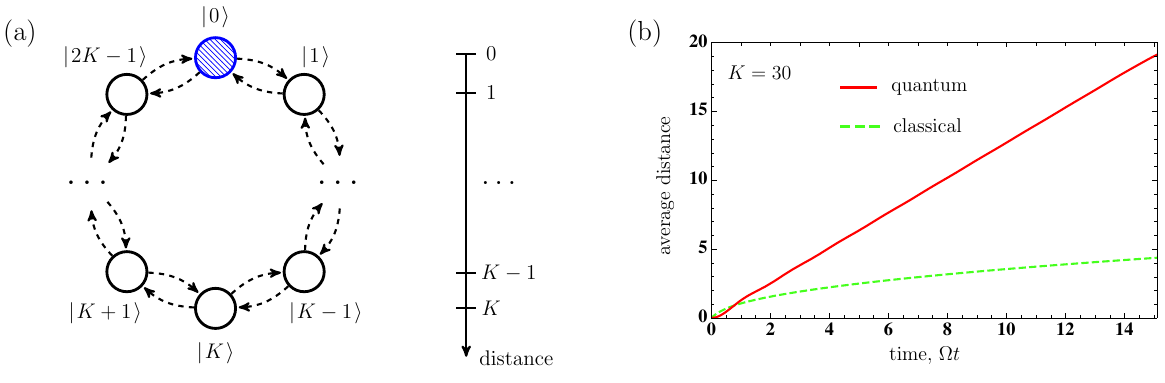}
   \end{center}
   \caption{(a) A cycle graph with $2K$ nodes. Each node represents a possible location of a particle. The particle is initially placed in node ``$0$'' and walks according to Eq.~(\ref{eq:singleEvolution}) or Eq.~(\ref{eq:classicalEvolution}) for classical and quantum particle, respectively. (b) A single quantum particle ``hits'' nodes at the average distance $\bar{l}$ quadratically faster. The simulations are performed on a cycle graph with $K=30$.}
   \label{fig:fig1}
   \end{figure}

Quantum walk on a cycle graph is schematically visualized in Fig.~\ref{fig:fig1}(a). The graph consists of nodes $v$ from the set $\mathcal{V}=\{0,\dots,2K-1\}$ and edges $e$ from the set $\mathcal{E}=\{(0,1),(1,2),\dots,(2K-2,2K-1),(2K-1,0)\}$. A quantum particle starts at the origin in the node $0$, and it's quantum state is fully described by it's position degree of freedom $\ket{\psi(0)}=\ket{0}$. This state is next evolved to some state $\ket{\psi(t)}=\sum_{v\in\mathcal{V}} \alpha_v(t) \ket{v}$ with probability amplitudes $\alpha_v$. The evolution of the quantum state is described by the Schr{\"o}dinger equation:
\begin{equation}
   	\ket{\psi(t)} = \mathrm{e}^{-\frac{i}{\hbar}\mathcal{H}t}\ket{0},
   \label{eq:singleEvolution}
\end{equation}
where a time-independent nearest-neighbor hopping Hamiltonian is
\begin{equation}
   	\mathcal{H} = \hbar~\frac{\Omega}{2}~\sum_{(v_1,v_2)\in\mathcal{E}} \sigma_{v_1}^x\sigma_{v_2}^x + \sigma_{v_1}^y\sigma_{v_2}^y,
   \label{eq:singleHamiltonian}
\end{equation}
with $\Omega$ being a tunneling rate, $\sigma^x$ and $\sigma^y$ are Pauli matrices acting on the node excitation degree of freedom.

In the case of classical random walks, a classical particle is described by a probability distribution $p(t)$ vector with $2K$ non-negative entries. The following differential equation describes the dynamics of this probability vector:
\begin{equation}
   	p(t) = \mathrm{e}^{(T-I)t}p_0,
   \label{eq:classicalEvolution}
\end{equation}
where $T$ is the transition matrix, and $I$ is the identity matrix. The transition matrix is the matrix of probabilities that are describing transitions on a graphs, and is closely related to the matrix in Eq.~(\ref{eq:singleHamiltonian}).

By placing a classical, or a quantum particle in the node $v=0$, we can ask a question about the time it will hit a vertex at a distance $l$ from $v=0$ (see Fig.~\ref{fig:fig1}(a), where the particle is marked with a filled circle). To estimate this time $t$ for the distance $l$, we calculate an average distance $\bar{l(t)}$. Fig.~\ref{fig:fig1}(b) shows the results of both classical and quantum walks simulations and the calculated $\bar{l(t)}$. From these simulations we can see that the quantum particle traverses the circle quadratically faster in average distance.\\\bigskip

It was observed, as e.g. shown by the example in Fig.~\ref{fig:fig1}(b), that quantum particles can propagate faster. However, how do two interacting quantum particles propagate on a graph compared to non-interacting quantum particles? What is the hitting time of two identical quantum particles? The answers are not clear, as interacting particles might repel from each other, and distribute uniformly on the cycle graph, without reaching the most distant nodes of the graph together.

\begin{figure}[t!]
   \begin{center}
   \includegraphics[width=0.45\linewidth]{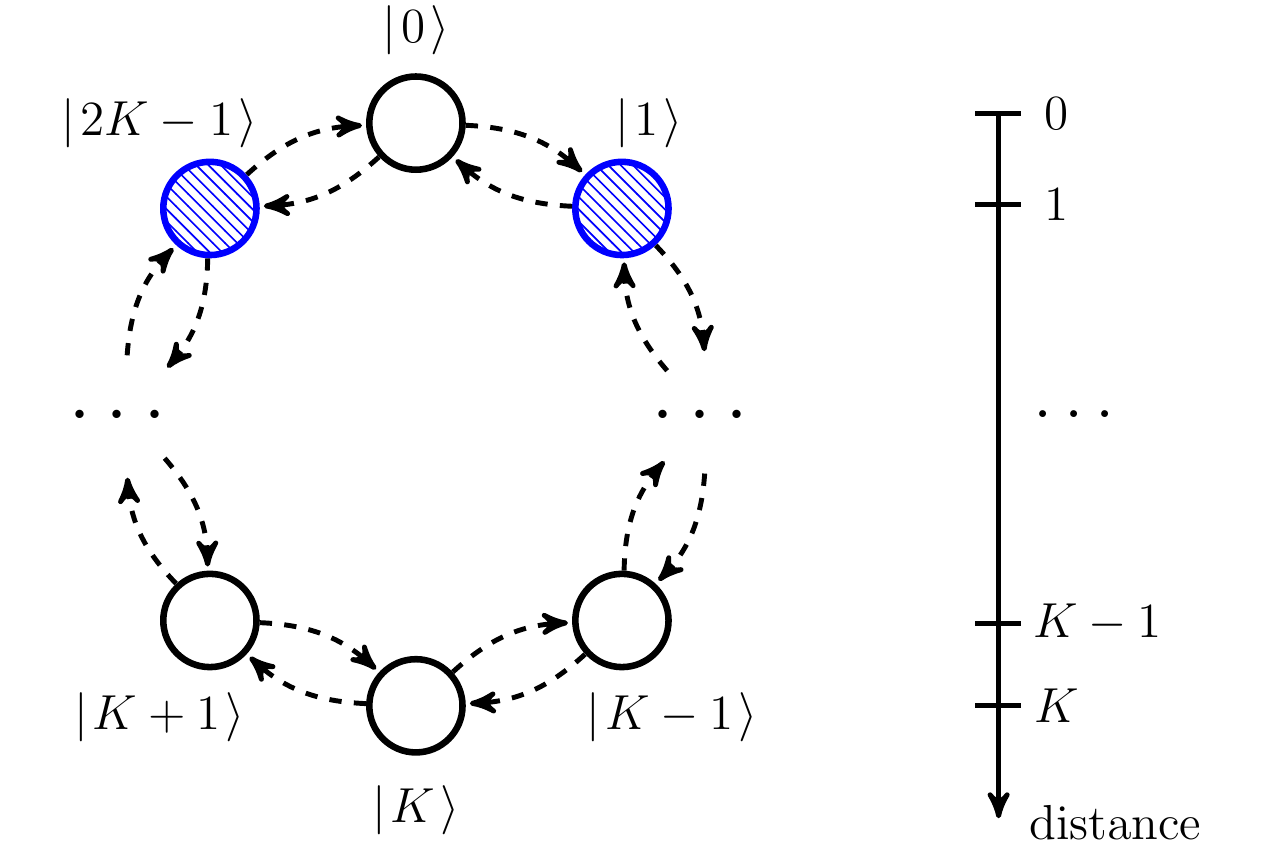}
   \end{center}
   \caption{(a) A cycle graph with $2K$ nodes. Each node represents a possible location of a particle. Two quantum particles are placed at distance $l=1$ from the origin.}
   \label{fig:fig2}
   \end{figure}

To answer these questions, we model two quantum particles that walk on a cycle graph, as shown in Fig.~\ref{fig:fig2}. As a potential physical implementation of the described quantum walks~\cite{wang2014physical} we consider electrons in semiconductor quantum dots. Physically, each node can be thought of as a semiconductor quantum dot and each edge as a barrier through which an electron can tunnel~\cite{fedichkin2000coherent,solenov2006continuous}. By this analogy, it is possible to represent a cycle graph as an array of tunnel-coupled quantum dots that are arranged in a circle. Technologies and techniques that are used to fabricate quadruple~\cite{takakura2014single,delbecq2014full} and quintuple~\cite{ito2016detection} quantum dots can be used for designing scalable architectures for realizing cycle graphs. The described physical implementation of quantum walks has several advantages for quantum computing. One of them is that semiconductor charge qubits can be protected from computational errors~\cite{fedichkin2004error,vasiliev2014estimations} with the help of standard quantum error correction algorithms~\cite{melnikov2013quantum,melnikov2013measure}, and by using a decoherence-free subspace qubit encoding~\cite{friesen2017decoherence}.

In a single electron regime of the described quantum system, the electron spreads coherently which corresponds to a quantum walk dynamics of a single particle~\cite{fedichkin2016quantum}, as in Fig.~\ref{fig:fig1}. In a two-electron regime particle can be treated as weakly interacting (or, non-interacting) and strongly interacting (or, interacting). In the non-interacting case, electrons are modeled as identical independent quantum particles, the evolution of which is described by Eq.~(\ref{eq:singleHamiltonian}). In the case of interacting particles, we consider two repelling electrons in the electric field of each other. This scenario is visualized schematically in Fig.~\ref{fig:fig2} and is described by the following quantum state~\cite{melnikov2016quantum}
\begin{equation}
   	\ket{\psi(t)} = \sum_{v_1=0}^{2K-2}\sum_{v_2=v_1+1}^{2K-1} \alpha_{v_1 v_2}(t) \ket{\psi_{v_1 v_2}}
   \label{eq:twoElectrons}
\end{equation}
where
\begin{equation}
   	\ket{\psi_{v_1 v_2}} = \frac{1}{\sqrt{2}}\left(\ket{v_1 v_2} - \ket{v_2 v_1}\right).
   \label{eq:twoElectronsPart}
\end{equation}
The evolution of the two interacting identical particles is governed by the hopping Hamiltonian with Coulomb-type approximation interaction term~\cite{melnikov2016quantum}:
\begin{align}
\mathcal{H} & = \hbar\Omega \sum_{v_1=0}^{2K-1}~\sum_{v_2=v_1+2}^{2K-3+v_1} \ket{(v_2+1)~\mathrm{mod}~2K,~v_1}\bra{v_2~\mathrm{mod}~2K,~v_1} + \ket{v_1,~(v_2+1)~\mathrm{mod}~2K}\bra{v_1,~v_2~\mathrm{mod}~2K}\nonumber\\
& + \ket{v_2~\mathrm{mod}~2K,~i}\bra{(v_2+1)~\mathrm{mod}~2K,~v_1} + \ket{v_1,~v_2~\mathrm{mod}~2K}\bra{v_1,~(v_2+1)~\mathrm{mod}~2K}.
    \label{eq:twoElectronsHamiltonian}
\end{align}
The detailed description of the origin and properties of Eq.~(\ref{eq:twoElectronsHamiltonian}) can be found in Ref.~\citenum{melnikov2017entanglement}. In both interacting and non-interacting cases we examine zero temperature limit avoiding influence of thermal effects on quantum transport, cf. Ref.~\citenum{lebedev2017exciton}.

   \begin{figure}[t!]
   \begin{center}
   \includegraphics[width=1\linewidth]{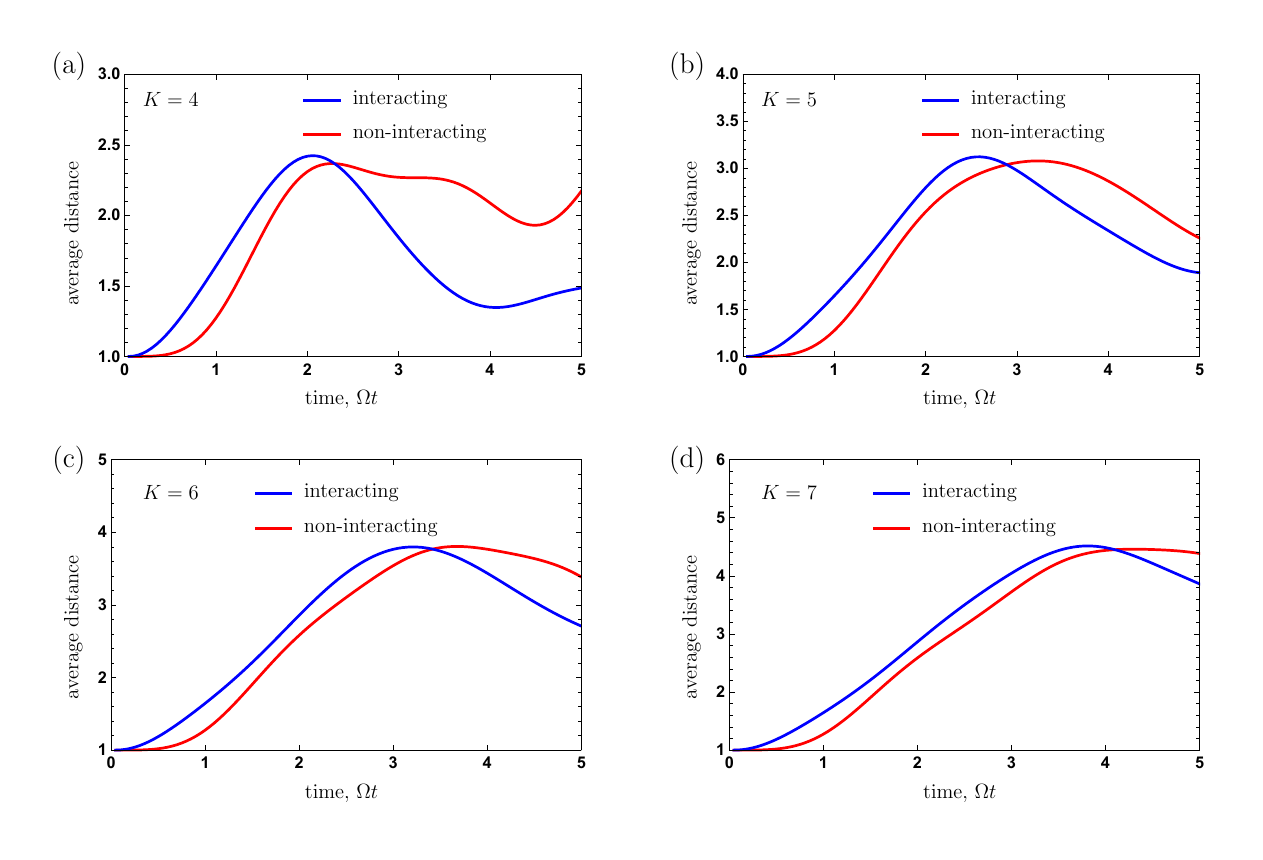}
   \end{center}
   \caption{The average propagation distance (defined in Eq.~(\ref{eq:averageDistance})) \textit{vs.} time. Interacting electrons propagate faster than non-interacting identical particles on cycle graphs of different sizes: $K=4$ ($8$ nodes, a), $K=5$ ($10$ nodes, b), $K=6$ ($12$ nodes, c), and $K=7$ ($14$ nodes, d).}
   \label{fig:fig3}
   \end{figure}

By simulating the time evolution of two particles, we calculate the average distance these particles cover after time $t$. By taking into account the anti-symmetric nature of the fermionic wave function, this distance is computed as
\begin{equation}
   	\overline{l(t)} = \sum_{l=1}^K l\rho^{(1)}_{l,l}(t) + \sum_{l=K+1}^{2K-1} (2K-l)\rho^{(1)}_{l,l}(t),
   \label{eq:averageDistance}
\end{equation}
with $\rho^{(1)}(t)$ being the reduced density matrix of the first system, which is also equal to $\rho^{(2)}(t)$ due to particle identity.

The results of calculating $\overline{l(t)}$ are shown in Fig.~\ref{fig:fig3}. The average distance for interacting particles is shown in blue, whereas the average distance for non-interacting particles is shown in red. It can be seen that interacting electrons propagate faster than non-interacting quantum particles. We observe that this happens consistently for all simulated graph sizes from $K=4$ (Fig.~\ref{fig:fig3}(a)) to $K=7$ (Fig.~\ref{fig:fig3}(d)). The difference between the interacting and non-interacting particles is especially visible during the initial time steps. To understand the reason behind this difference, we look at correlations in positions of interacting particles. The quantum correlations between electrons, or fermionic entanglement, can be quantified by the fermionic concurrence~\cite{majtey2016multipartite}.

   \begin{figure}[t!]
   \begin{center}
   \includegraphics[width=1\linewidth]{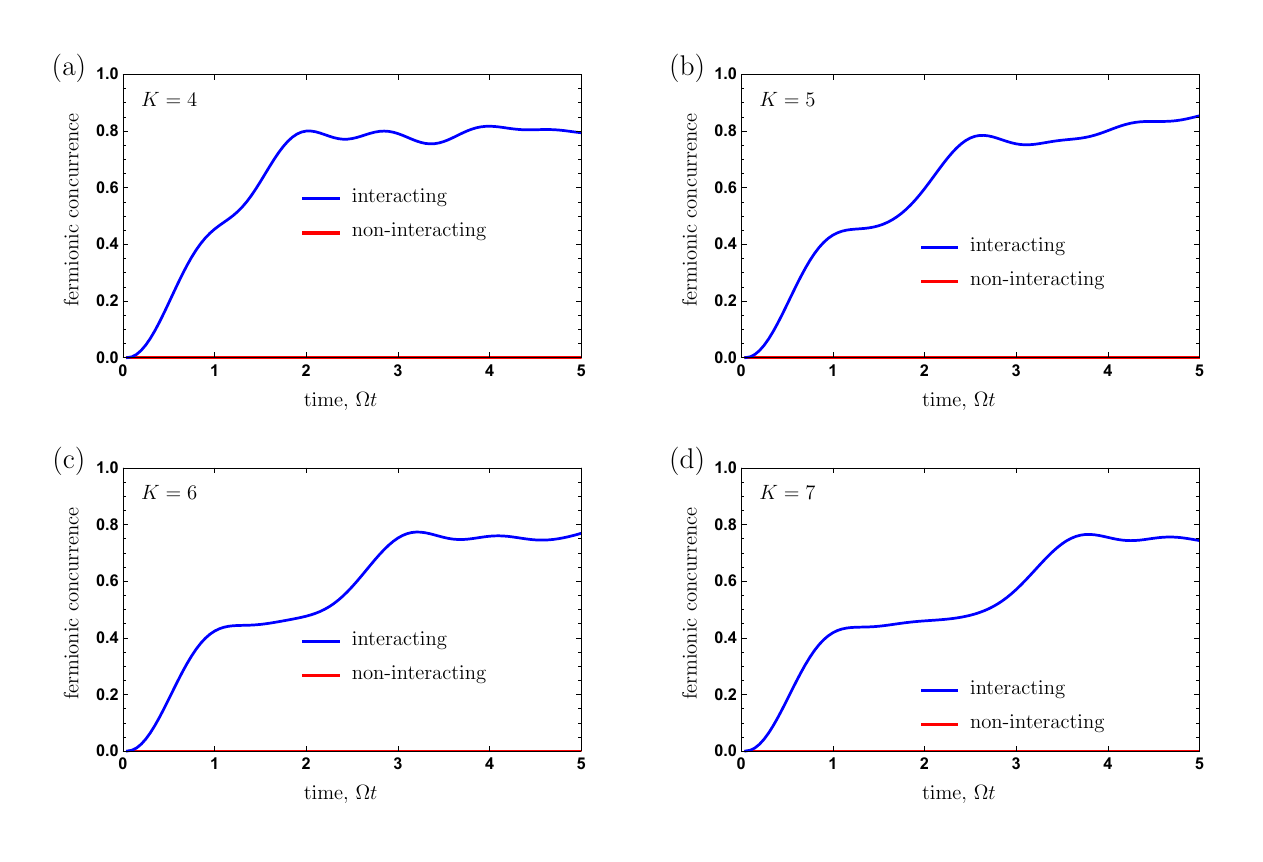}
   \end{center}
   \caption{The fermionic concurrence for different cycle graph sizes: $K=4$ ($8$ nodes, a), $K=5$ ($10$ nodes, b), $K=6$ ($12$ nodes, c), and $K=7$ ($14$ nodes, d).}
   \label{fig:fig4}
   \end{figure}

Fig.~\ref{fig:fig4} shows how the fermionic entanglement evolves over time for graphs of sizes $K=4$ (a), $K=5$ (b), $K=6$ (c), and $K=7$ (d). If quantum particles do not interact with each other (red lines), no entanglement is generated. From Fig.~\ref{fig:fig4} one can see that similar to the case of the average traversed distance, the entanglement of interacting particles grows up to a certain point. However, after this point, the fermionic entanglement does not go down (as the average distance does), but saturates. A more comprehensive analysis of the shown entanglement dynamics of interacting fermions can be found in Refs.~\citenum{melnikov2017entanglement,melnikov2018fermionic}, where realistic noise scenarios a taken into account (this includes a depolarization noise that can dissociate and annihilate entanglement of many particles~\cite{filippov2013dissociation}).

   \begin{figure}[t!]
   \begin{center}
   \includegraphics[width=1\linewidth]{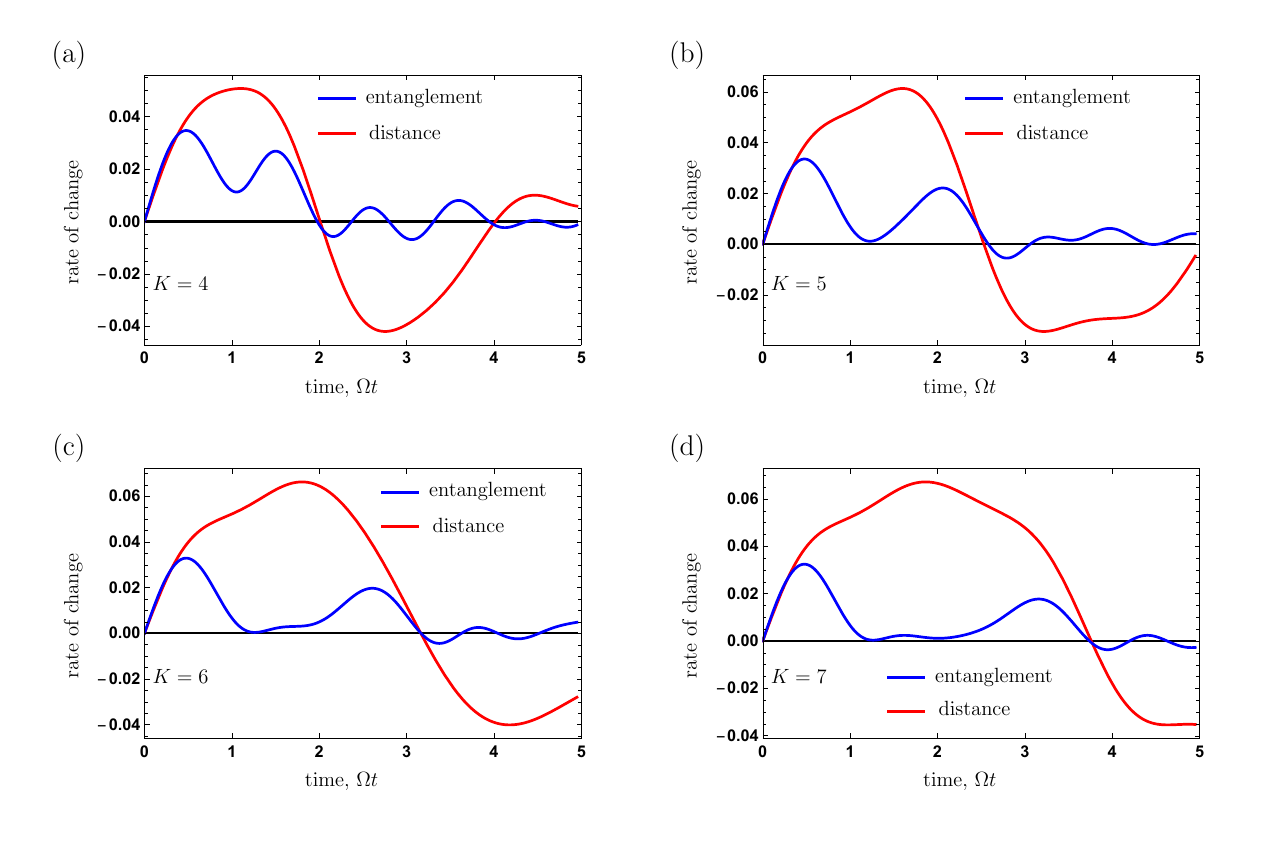}
   \end{center}
   \caption{The rate of change in fermionic entanglement (blue) and average distance (red) for interacting fermions for different cycle graph sizes: $K=4$ ($8$ nodes, a), $K=5$ ($10$ nodes, b), $K=6$ ($12$ nodes, c), and $K=7$ ($14$ nodes, d).}
   \label{fig:fig5}
   \end{figure}

To further understand the correlations between average distance and fermionic entanglement, we compute the rate (derivatives) of change in average distance and fermionic entanglement. These rates are shown in Fig.~\ref{fig:fig5}(a)-(d). The results demonstrate that the rate of change in distance (red) and entanglement (blue) is almost the same at the very beginning. This highlights the fact that both fermionic entanglement and hitting time are related.
Importantly, both red and blue curves are positive during the same time interval, and they go negative almost at the same time. By this, we observe that the interacting particles spread further in the direction of the opposite node as long as entanglement grows.

\section{Conclusion}
\label{sec:conclusion}

In this paper, we studied continuous-time quantum walk dynamics of two indistinguishable particles in a cycle graph. The quantum walk dynamics of this system can lead to entanglement given some physical interaction between particles, e.g., mutual repulsion. These dynamics were studied before and shown to be useful for the preparation of two-qudit entangled states of two distinguishable subsystems~\cite{melnikov2016quantum,melnikov2016continuous,melnikov2017entanglement,melnikov2018fermionic}. The highly entangled states of qudits can be obtained by only using the free quantum evolution of identical particles, without relying on any additional manipulations with particles.

Here we studied a characteristic property of quantum walk dynamics -- the hitting time of these indistinguishable particles -- and related this property to fermionic entanglement.
We demonstrated that interacting electrons walk on cycle graphs hit distant nodes faster than non-interacting particles. This speed-up is also reflected in increasing entanglement as particles propagate further. Finally, the maximum propagation distance is shown to correspond to the maximum of fermionic entanglement.
Our results emphasize the role and importance of fermionic entanglement in quantum transport.

\acknowledgments 

This work was financially supported by Government of Russian Federation, Grant 08-08 and by Ministry of Education and Science of the Russian Federation within the Federal Program ``Research and development in priority areas for the development of the scientific and technological complex of Russia for 2014-2020'', Activity 1.1, Agreement on Grant No. 14.572.21.0008 of 23 October, 2017, unique identifier: RFMEFI57217X0008.


\bibliographystyle{spiebib} 

\end{document}